\documentstyle[aps,twocolumn,epsf]{revtex}
\begin{document}
\draft
\title{An extreme critical space-time: echoing and black-hole perturbations}
\author{Sean A. Hayward}
\address{Center for Gravitational Physics and Geometry,
104 Davey Laboratory, The Pennsylvania State University,
University Park, PA 16802-6300, U.S.A.\\
{\tt hayward@gravity.phys.psu.edu}}
\date{12th April 2000}
\maketitle
\begin{abstract}
A homothetic, static, spherically symmetric solution 
to the massless Einstein-Klein-Gordon equations is described.
There is a curvature singularity 
which is central, null, bifurcate, massless and marginally trapped.
The space-time is therefore extreme in the sense of 
lying at the threshold between black holes and naked singularities, 
just avoiding both.
A linear perturbation analysis reveals two types of dominant mode.
One breaks the continuous self-similarity 
by periodic terms reminiscent of discrete self-similarity,
with echoing period within a few percent of the value observed numerically 
in near-critical gravitational collapse.
The other dominant mode explicitly produces a black hole, 
white hole, eternally naked singularity or regular dispersal,
the latter indicating that the background is critical.
The black hole is not static but has constant area,
the corresponding mass being linear in the perturbation amplitudes,
explicitly determining a unit critical exponent.
It is argued that 
a central null singularity may be a feature of critical gravitational collapse.
\end{abstract}
\pacs{04.20.Dw, 04.70.Bw, 04.20.Jb, 04.25.Nx}

\section{Introduction}

The discovery by Choptuik\cite{C1} of critical phenomena 
in gravitational collapse
is widely regarded as the most unexpected success of numerical relativity.
The now famous mass scaling law and echoing behavior
had not been anticipated by existing black-hole theory.
Despite a subsequent theoretical framework involving dynamical-systems 
and renormalization-group methods\cite{KHA,HKA},
this is still not fully understood;
see reviews by Choptuik\cite{C2} and Gundlach\cite{G1}.
In particular, the exactly critical space-time 
suggested by such numerical simulations is not known analytically.
Thus it is of some interest to study solutions with similar properties.

The numerical results suggest a self-similar critical space-time.
For the massless Einstein-Klein-Gordon case originally studied by Choptuik,
there is a discrete self-similarity 
with a characteristic echoing period\cite{C1,G2}.
Other matter models yield continuous self-similarity,
which can be described by a homothetic vector.
A class of homothetic, spherically symmetric solutions
to the massless Einstein-Klein-Gordon equations was given by Roberts\cite{R} 
and has been studied by various authors\cite{B,ONT,F1,F2}.
This article describes another such solution, which is static.
This solution, the Roberts class and its time-inverse form 
a two-parameter family which was actually given by Brady\cite{B},
though in that reference a parameter was fixed 
to reduce to the Roberts class.

Since these solutions are continuously rather than discretely self-similar,
they do not include the critical space-time itself.
However, one may hope to find a perturbed space-time 
close to the critical space-time.
Indeed, Frolov\cite{F2} has argued that 
linear perturbations of the critical Roberts solution
break the self-similarity from continuous to discrete.
This article therefore studies the static solution 
and its spherically symmetric linear perturbations.
There are indeed perturbations which break the continuous self-similarity
to a certain periodic structure,
actually falling short of discrete self-similarity
for reasons explained subsequently.
The static symmetry greatly simplifies the analysis.
In particular, 
this allows the first explicit construction of perturbed black holes 
with finite mass from such a critically trapped space-time.

The article is organised as follows.
Section~II gives the solution 
and describes the global structure of the space-time.
Section~III studies linear perturbations, 
explicitly giving the general solution to the linearized field equations,
then imposing appropriate boundary conditions and finding the relevant modes.
Section~IV describes a type of dominant mode which produces echoing.
Section~V describes the other dominant mode, 
which produces black holes for a certain quadrant in amplitude space.
Section~VI shows that the other quadrants produce white holes, 
naked singularities and regular dispersal respectively.
Section~VII concludes with some speculations concerning critical collapse.

\section{The solution}

\begin{figure}
\centerline{\epsfxsize=8cm \epsfbox{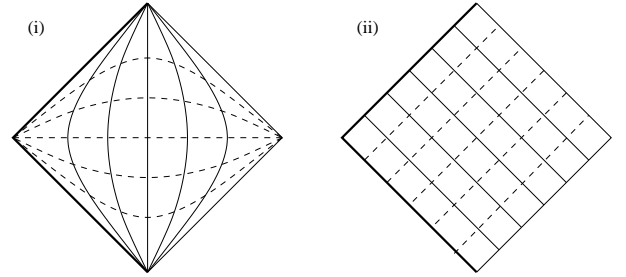}}
\caption{Penrose diagrams of the space-time.
The thick line is the central ($r=0$) singularity 
and the other conformal boundary is null infinity ($r=\infty$).
(i) The dashed lines are the static hypersurfaces ($\tau$ constant) and 
the solid lines are the orthogonal homothetic hypersurfaces ($\rho$ constant).
(ii) The dashed lines are of constant $x^-$ or $\xi^-$
and the solid lines are of constant $x^+$ or $\xi^+$,
which are also homothetic hypersurfaces.}
\label{penrose}
\end{figure}

The field equations for Einstein gravity 
with a massless Klein-Gordon field $\phi$,
fixing the relative coupling constant, are
\begin{eqnarray}
&&R=2\nabla\phi\otimes\nabla\phi\\
&&\nabla^2\phi=0
\end{eqnarray}
where $R$ is the Ricci tensor and $\nabla$ the covariant derivative 
of the metric $g$.
A spherically symmetric metric may be written locally 
in terms of the line-element
\begin{equation}
ds^2=r^2d\Omega^2-2e^{2\gamma}dx^+dx^-
\end{equation}
where $d\Omega^2$ is the line-element of the unit sphere
and $(r,\gamma)$ are functions of the null coordinates $x^\pm$.
Then the area of the spheres is $4\pi r^2$, so that $r$ is the areal radius.
The field equations in these coordinates are\cite{sph}
\begin{eqnarray}
&&\partial_\pm\partial_\pm r-2\partial_\pm\gamma\partial_\pm r
=-r(\partial_\pm\phi)^2\\
&&r\partial_+\partial_-r+\partial_+r\partial_-r+e^{2\gamma}/2=0\\
&&r^2\partial_+\partial_-\gamma-\partial_+r\partial_-r-e^{2\gamma}/2
=-r^2\partial_+\phi\partial_-\phi\\
&&r\partial_+\partial_-\phi+\partial_+r\partial_-\phi+\partial_-r\partial_+\phi
=0.
\end{eqnarray}
It will also be convenient to use coordinates $(\rho,\tau)$ defined locally by
\begin{equation}
x^\pm=\pm e^{\rho\pm\tau}.
\end{equation}
The solution is given simply by
\begin{equation}
r=e^\rho\qquad\gamma=0\qquad\phi=\tau
\end{equation}
which is the case $\alpha=\beta=0$ of Brady\cite{B}.
The variables $(X,Y)$ of Choptuik\cite{C1,C2} evaluate as $(0,1/2)$.
The line-element can be written as
\begin{equation}
ds^2=e^{2\rho}(d\Omega^2+2d\rho^2-2d\tau^2).
\end{equation}
This is explicitly conformal to a flat metric times a unit sphere, 
so that a Penrose diagram can be given as in Fig.\ref{penrose}.
The metric is evidently static, 
with static Killing vector $\partial/\partial\tau$.
It is also homothetic, or continuously self-similar, 
with homothetic vector $\partial/\partial\rho$.
This means that
\begin{equation}
L_{\partial/\partial\rho}g=2g
\end{equation}
where $L$ denotes the Lie derivative and the factor of 2 is conventional.
It follows that the null vectors $\partial/\partial\xi^\pm$ 
are also homothetic vectors, where
\begin{equation}
\xi^\pm=\pm\ln(\pm x^\pm)=\tau\pm\rho.
\end{equation}
In this sense the solution is doubly self-similar.
Hypersurfaces orthogonal to a homothetic vector may be called 
homothetic hypersurfaces.

There is a central singularity, meaning $r=0$ there, at $x^+x^-=0$.
This is a bifurcate null singularity, 
where the causal nature is defined by the conformal metric.
It is therefore also marginally trapped, since $\partial_\pm r=0$ on $x^\mp=0$.
It is a curvature singularity, 
since the Ricci scalar is found to be just $-r^{-2}$.
The other conformal boundary corresponds to infinite $r$
and constitutes null infinity, though the space-time is not asymptotically flat.
Correspondingly, the mass or energy\cite{sph,1st}
\begin{equation}
m=(r/2)(1-g^{-1}(\nabla r,\nabla r))
\label{mass}
\end{equation}
evaluates as $m=r/4$, becoming infinite at null infinity.
The singularity is massless in the sense that $m=0$ there.
This confirms the view that 
massless central singularities lie at the threshold between 
generic trapped and naked singularities,
which occur for $m>0$ and $m<0$ respectively\cite{sph}.
The space-time is summarized in Fig.\ref{penrose}.
The $(\rho,\tau)$ or $\xi^\pm$ coordinates take all real values,
whereas the $x^\pm$ coordinates have the range $x^+>0$, $x^-<0$.
The future and past branches of the singularity 
occur for $x^-=0$ and $x^+=0$ respectively,
where $\partial/\partial\tau$ is taken to be future-pointing.

The solution just avoids being a black hole (or white hole) in the sense that
there are no trapped surfaces, yet the singularity is marginally trapped.
Also, the singularity just avoids being naked in the sense of Penrose\cite{P},
that it should lie to the future of some point and the past of another point 
on some observer's worldline. 
Likewise, an observer could see arbitrarily high but not infinite curvature 
from the future branch of the singularity.
In summary, this is a new type of extreme space-time, 
lying at the threshold between black holes and naked singularities.

\section{Linear perturbations}

Henceforth, 
spherically symmetric linear perturbations of the space-time will be studied,
initially following the analysis of the critical Roberts solution 
by Frolov\cite{F1,F2}.
The linear perturbations studied here may be decomposed 
by Laplace transformations into modes
\begin{eqnarray}
&&r=e^\rho(1+\epsilon\tilde r(\tau)e^{-k\rho})\\
&&\gamma=\epsilon\tilde\gamma(\tau)e^{-k\rho}\\
&&\phi=\tau+\epsilon\tilde\phi(\tau)e^{-k\rho}
\end{eqnarray}
governed by the complex frequency $k$.
Assuming this form of the perturbations fixes the perturbative gauge freedom,
apart from the zero point of $\tau$.
The field equations are linearized 
by removing terms $o(\epsilon)$ in the perturbation parameter, $0<\epsilon\ll1$.
The four independent components of the field equations yield 
three independent linearized equations:
\begin{eqnarray}
&&\tilde\gamma=(1-k/2)^2\tilde r-\tilde r''/4\\
&&\tilde\phi=-\tilde r'-\tilde\gamma'/k\\
&&\tilde r''+k^2\tilde r+2k\tilde\gamma+2\tilde\phi'=0
\end{eqnarray}
where $f'=\partial f/\partial\tau$.
The first two are solved for $(\tilde\gamma,\tilde\phi)$ explicitly 
if $\tilde r$ is known, 
so that the system reduces to a single fourth-order equation for $\tilde r$:
\begin{equation}
\tilde r''''-2(k^2-k+2)\tilde r''+k^2(k^2-2k+4)=0.
\end{equation}
This linear ordinary differential equation can be solved 
via the auxiliary equation
\begin{equation}
\omega^4-2(k^2-k+2)\omega^2+k^2(k^2-2k+4)=0.
\end{equation}
This is a biquadratic equation, so that the roots come in pairs,
$\omega=\pm\omega_A$, $\pm\omega_B$ where
\begin{eqnarray}
&&\omega_A=(k^2-2k+4)^{1/2}\\
&&\omega_B=k.
\end{eqnarray}
Then the general solution for each $k$ is
\begin{equation}
\tilde r=A_+e^{\omega_A\tau}+A_-e^{-\omega_A\tau}
+B_+e^{\omega_B\tau}+B_-e^{-\omega_B\tau}
\end{equation}
for constants $A_\pm$, $B_\pm$,
except in special cases where roots coincide, 
when similarly standard solutions can be given.
The special cases are (i) $k=0$, when $\omega_B=0$,
(ii) $k=2$, when $\omega_A=\omega_B$,
and (iii) $k=1\pm\sqrt{3}i$, when $\omega_A=0$.

\begin{figure}
\centerline{\epsfxsize=3cm \epsfbox{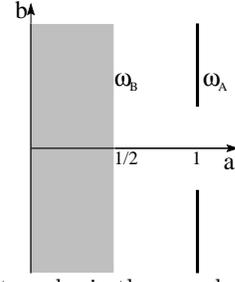}}
\caption{Relevant modes in the complex plane of $k=a+ib$.
The $\omega_B$ modes fill a region
while the $\omega_A$ modes form two lines.}
\label{modes}
\end{figure}

Weak boundary conditions will be taken as
\begin{equation}
\lim_{x^\pm\to\pm\infty}r\not=0\qquad
\lim_{x^\pm\to0}r^{-1}\not=0.
\label{weak}
\end{equation}
If these conditions were not met, 
the perturbed $r$ would vanish at the original infinity 
or become infinite at the original centre, respectively, 
either of which would be wildly different from the background space-time
and therefore inconsistent with the linear approximation.
Slightly stronger boundary conditions additionally impose
\begin{equation}
\exists\lim_{x^\pm\to\pm\infty}r^{-1}\qquad
\exists\lim_{x^\pm\to0}r
\label{strong}
\end{equation}
with similar justification.
By writing a mode explicitly as
\begin{eqnarray}
r&=&e^\rho(1+\epsilon Ae^{\omega\tau-k\rho})\nonumber\\
&=&(-x^+x^-)^{1/2}+\epsilon A(x^+)^{(1-k+\omega)/2}(-x^-)^{(1-k-\omega)/2}
\label{mode}
\end{eqnarray}
the weak boundary conditions can be seen to imply
\begin{equation}
\Re(k\pm\omega)\le1
\end{equation}
with cases of equality examined individually in the following.
For the $\omega_B$ modes this means
\begin{equation}
a\le1/2
\end{equation}
where $a=\Re k$, $b=\Im k$, 
whereas for the $\omega_A$ modes a little calculation yields
\begin{equation}
a=1\qquad b^2\ge3
\end{equation}
with $\omega_A=i\beta$, $\beta^2=b^2-3$.
These modes are indicated in Fig.\ref{modes}.

\section{Echoing}

A common argument is that the relevant modes are those with $a>0$
and the dominant modes those with maximal $a$\cite{KHA,HKA,C2,G1,G2,F1,F2}.
As depicted in Fig.\ref{modes}, this occurs for the $\omega_A$ modes.
One might also expect the dominant mode among these to be 
that with minimal $|b|$, since higher-frequency modes interfere destructively.
This mode is given by $k=1\pm\sqrt{3}i$, case (iii) above:
\begin{equation}
r=e^\rho+\epsilon A\cos(\sqrt{3}\rho+c).
\end{equation}
The general solution for this mode would replace $A$ with $A+C\tau$,
but if $C\not=0$ the boundary conditions (\ref{weak}) are not satisfied.
The perturbed space-time is static and the perturbation $\delta r$ is periodic,
with echoing in $\rho$ of period 
\begin{equation}
\Delta=2\pi/\sqrt{3}=3.62\ldots
\end{equation}
which is within about $5\%$ of the value 3.44 
obtained numerically by Choptuik\cite{C1}.
This suggests that 
the perturbed space-time may be close to the critical space-time.
However, the Klein-Gordon perturbation $\delta\phi$ vanishes for this mode.

Note that the perturbed space-time is not discretely self-similar;
despite $\delta r$ being periodic in $\rho$, 
$r$ contains an exponential term in $\rho$.
The situation may be understood schematically in terms of a phase-space picture.
The universality of the critical exponent suggests that 
the critical solution is an intermediate attractor 
of codimension one\cite{KHA,HKA,C2,G1}.
In other words, there is a critical hypersurface in phase space,
separating black-hole solutions from dispersal solutions.
The background solution lies in this hypersurface,
by the same argument as for the critical Roberts solution: 
the Roberts class, 
cut and pasted to flat regions as described by Brady\cite{B},
describe either black-hole formation or dispersal, 
except for the delimiting critical case.
Linear perturbations are represented by vectors from the background solution.
One seeks a perturbation tangent to the critical hypersurface, 
pointing towards the critical space-time.
If the critical hypersurface is curved,
the perturbed solution will generally be outside the hypersurface 
and so exhibit exponential as well as oscillatory behavior.

For the modes with $k=1+ib$, $b^2>3$, one finds
\begin{equation}
r=e^\rho+\epsilon A\cos(b\rho+c)\cos(\beta\tau+d)
\end{equation}
so that the perturbations are periodic in both $\rho$ and $\tau$.
The variable $X$ plotted by Choptuik\cite{C1,C2} is, to $O(\epsilon)$, 
\begin{equation}
{1\over2}{\partial\phi\over{\partial\rho}}
=-{\epsilon Ab\beta\over4}\sin(b\rho+c)\sin(\beta\tau+d)
\end{equation}
which is explicitly sinusoidal in both $\rho$ and $\tau$.
Apart from the well known echoing in $\tau-\rho$,
roughly sinusoidal oscillations in $\rho$ at constant $\tau$ 
do seem to develop in the numerical simulations, 
e.g.\ figure 3 of Choptuik\cite{C2}.

\section{Black holes}

The common argument\cite{KHA,HKA,C2,G1,G2,F1,F2} that 
the dominant mode will occur for maximal $\Re k$, 
as this is largest as $\rho\to-\infty$,
implicitly assumes that the other coordinate $\tau$ is finite,
which actually applies only where the singularity bifurcates.
For the rest of the singularity, 
the explicit expression (\ref{mode}) shows that 
the dominant mode at $x^\mp=0$ instead occurs for maximal $\Re(k\pm\omega)$.
As above, this maximal value is unity.
The corresponding modes consist of those discussed in the previous section
and the $\omega_B$ modes with $\Re k=1/2$.
Non-linear perturbations may then be dominated by either or both,
in a way which is unclear from a linear perturbation analysis.
Henceforth the new modes will be discussed separately.

The $\omega_B$ modes with $k=1/2+ib$, $b\not=0$,
do not satisfy the boundary conditions (\ref{strong}),
since $r$ does not have a limit as $x^\pm\to0$, $x^\mp\not=0$.
Thus the dominant mode is just $\omega_B=k=1/2$.
Then
\begin{eqnarray}
r&=&e^\rho
+\epsilon B_+e^{(\rho+\tau)/2}+\epsilon B_-e^{(\rho-\tau)/2}\nonumber\\
&=&(-x^+x^-)^{1/2}+\epsilon B_+(x^+)^{1/2}+\epsilon B_-(-x^-)^{1/2}.
\end{eqnarray}
The locus of the centre $r=0$ can be determined by noting that 
it is a hyperbola in $z^\pm=(\pm x^\pm)^{1/2}$ coordinates:
\begin{equation}
0=z^+z^-+\epsilon B_+z^++\epsilon B_-z^-.
\end{equation}
Likewise, the locus of the trapping horizon, if any, 
is determined by
\begin{equation}
\partial_\pm r=\pm(\pm x^\pm)^{-1/2}((\mp x^\mp)^{1/2}+\epsilon B_\pm)/2
\end{equation}
where $\partial_\pm=\partial/\partial x^\pm$.
Here a trapping horizon\cite{1st,bhd} is a hypersurface 
foliated by marginal surfaces, $\partial_+r=0$ or $\partial_-r=0$,
which respectively characterize black or white holes.
These loci can be seen to be straight lines 
of constant $x^-$ or $x^+$ respectively, or non-existent,
depending on the signs of $B_\pm$.
These lines correspond to the asymptotes 
\begin{equation}
z^\mp=-\epsilon B_\pm
\end{equation}
of the above hyperbola,
which passes through $z^+=z^-=0$.
The part of the hyperbola and its asymptotes inside $z^+>0$, $z^->0$
can be mapped back to the $x^\pm$ coordinates,
the loci depending qualitatively only on the signs of $B_\pm$.
This, combined with the nature of the background conformal boundaries, 
determines the qualitative global structure of the perturbed space-time.

\begin{figure}
\centerline{\epsfxsize=8cm \epsfbox{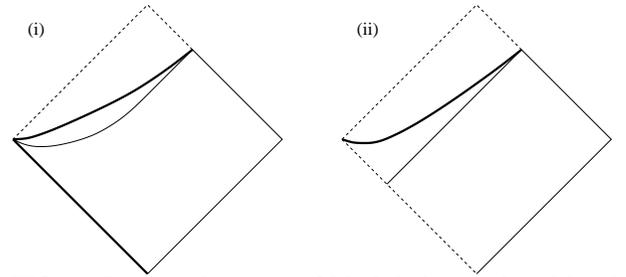}}
\caption{Penrose diagrams of black holes 
produced by the $\omega_B$ modes with $b=0$, 
for (i) $0<a<1/2$ and (ii) the dominant mode $a=1/2$.
The thick curve is the central singularity 
and the other curve is the trapping horizon,
between which lies the trapped region.
This occurs for the quadrant $B_+<0$, $B_->0$. 
Dashed lines indicate conformal boundaries of the background space-time 
which are either outside or inside the perturbed space-time.}
\label{bhs}
\end{figure}

For the quadrant $B_+<0$, $B_->0$, assumed in the remainder of this section,
there is a black hole,
with a region of trapped surfaces following a trapping horizon 
and preceding a spatial central singularity, as depicted in Fig.\ref{bhs}.
A mode with $b=0$, $0<a<1/2$ is also depicted for comparison,
the analysis being similar.
In that case, the trapping horizon is spatial, appears at zero area 
and evolves towards infinite area at conformal infinity.
However, for the dominant mode, 
a remarkable feature is that the trapping horizon is null, 
occuring at constant $x^-$, with constant areal radius
\begin{equation}
r=-\epsilon^2B_+B_-.
\end{equation}
This seems to be an example of an isolated horizon\cite{ABF}
which is not a Killing horizon; the perturbed space-time is not static.
It is also clearly an event horizon,
generalized from the usual definition for asymptotically flat space-times.

A curious feature is that the past branch of the background singularity 
is regularized by this perturbation:
the $x^\pm$ coordinates break down at $x^+=0$,
but it can be shown to have finite area, finite mass $m=r/2$ 
and vanishing Ricci scalar.
(Essentially, $\gamma\to\infty$ there, 
so norms $\nabla f\cdot\nabla f=-2e^{-2\gamma}\partial_+f\partial_-f$
tend to vanish, the exponential term suppressing negative powers).
This implies that it is a degenerate trapping horizon\cite{1st,bhd}.
Extension inside this past event horizon seems, however, 
beyond the validity of the linear approximation;
it is meaningless to have a linear perturbation of a region 
which does not exist in the background space-time.

Since the mass or energy (\ref{mass}) is $m=r/2$ on a trapping horizon,
there is a mass scaling law
\begin{equation}
m=-\epsilon^2B_+B_-/2.
\end{equation}
Thus the mass is explicitly linear in the perturbation amplitudes 
$|\epsilon B_\pm|$ of both ingoing and outgoing modes,
so that the critical exponent is unity.
The renormalization-group argument\cite{KHA,HKA,C2,G1,G2,F1,F2},
that the critical exponent is the reciprocal of the maximal $\Re k$,
would not give this value unless $\Re k$ is replaced by $\Re(k+\omega)$, 
in accordance with the argument at the beginning of this section.
This reflects the double self-similarity of the solution:
the relevant homothetic vector at $x^\mp=0$ is $\pm\partial/\partial\xi^\pm$,
rather than $\partial/\partial\rho$.

\section{Dispersal, naked singularities and white holes}

\begin{figure}
\centerline{\epsfxsize=8cm \epsfbox{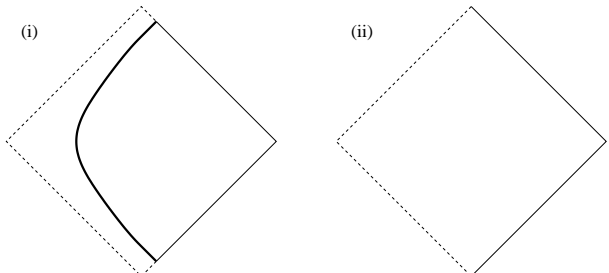}}
\caption{Penrose diagrams of space-times 
produced by the dominant $\omega_B$ mode, $k=1/2$, 
as for Fig.\ref{bhs}(ii).
(i) The quadrant $B_+<0$, $B_-<0$,
producing an eternally naked central singularity.
(ii) The quadrant $B_+>0$, $B_->0$, describing regular dispersal.}
\label{ns}
\end{figure}

The other quadrants of the dominant $\omega_B$ mode, $k=1/2$,
may be similarly analysed.
The quadrant $B_+>0$, $B_-<0$ is a time-reverse of the black-hole case,
yielding a white hole.
For the quadrant $B_+<0$, $B_-<0$, 
there is an eternally naked central singularity, 
with no trapped region outside the singularity.
For the quadrant $B_+>0$, $B_->0$, 
both branches of the background singularity are regularized:
the area vanishes only at $x^+=x^-=0$,
$x^\pm=0$ otherwise having finite area, finite mass $m=r/2$ 
and vanishing Ricci scalar, by a similar argument as for the black-hole case.
Extension beyond these horizons again seems
beyond the validity of the linear approximation.
As far as can be judged from the unextended region,
there is no singularity or trapped region,
with the perturbed solution describing regular dispersal.
The dispersal and naked-singularity solutions are depicted in Fig.\ref{ns},
the corresponding white-hole diagram being a time-reverse of Fig.\ref{bhs}(ii).

\begin{figure}
\centerline{\epsfxsize=4cm \epsfbox{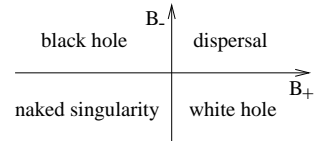}}
\caption{Summary of the effects of the dominant $\omega_B$ mode, $k=1/2$, 
in terms of the amplitudes $B_\pm$.
The background space-time lies at the intersection.}
\label{quad}
\end{figure}

This indicates that, in the phase-space picture,
the background space-time lies not only in the critical hypersurface 
separating black-hole formation from dispersal,
but also in another hypersurface separating black-hole formation
from naked singularities, 
indicating a fourth region presumably describing white-hole decay.
The picture would be analogous to Fig.\ref{quad},
with the critical space-time presumably lying roughly in the direction 
analogous to a third axis $A$ for the dominant $\omega_A$ mode.

\section{Remarks}

The solution studied here describes a new type of extreme space-time,
just on the verge of containing a black hole, white hole or naked singularity,
without actually containing any of them.
Each can be explicitly constructed by linear perturbations,
as can regular dispersal solutions,
so that the space-time is also critical;
it lies at the threshold between black holes, 
white holes, naked singularities and dispersal.
The symmetry and simplicity of the solution suggest it as a standard example 
of an extreme critical space-time.
As such, it may be useful in the context of cosmic censorship\cite{P,K}.
It should be noted that none of the perturbations studied here 
indicates a violation of cosmic censorship;
the naked-singularity perturbations involve an eternally naked singularity,
not one forming from regular initial conditions.

Various properties of the solution or its perturbations are analogous 
to those of near-critical numerical solutions in gravitational collapse.
In particular, the dominant black-hole perturbation produces a black hole 
of constant mass and area,
explicitly related to the perturbation amplitudes by a scaling law.
Other perturbations produce echoing reminiscent of the numerical solutions.

It may be conjectured that 
the numerically suggested critical space-time has a similar global structure
to that of the solution described here,
though this contradicts the analysis of Gundlach\cite{G2}.
Of course,
the global structure of a near-critical numerical solution is quite different,
asymptotically flat with an initially regular centre.
The critical and near-critical space-times should agree only
close to the future branch of the singularity.
A null central singularity in spherical symmetry is marginally trapped
and therefore lies at the threshold of black-hole or white-hole formation,
for the future-null and past-null cases respectively.
An observer close to a future-null singularity could see 
arbitrarily high but not infinite curvature, 
consistently with the numerical results.
Likewise, a future-null singularity is not naked,
in the sense that there is no future-causal curve from it,
but is just on the verge of nakedness; 
to coin a phrase, a teasing singularity.

\begin{figure}
\centerline{\epsfxsize=8cm \epsfbox{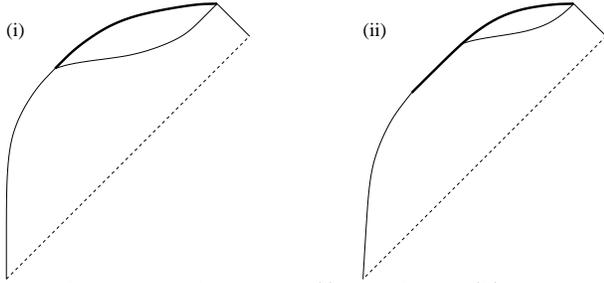}}
\caption{Penrose diagrams of (i) generic and (ii) non-generic black holes 
formed from initial data on an outgoing null hypersurface (dashed line),
according to the analysis of Christodoulou.
The initially regular centre becomes a central singularity (thick curve)
with the trapping horizon (the other curve) emerging 
as the singularity becomes spatial. 
In the generic case (i) the centre is instantaneously null,
while in the non-generic case (ii) it is null for a finite time,
during which it coincides with the trapping horizon in a conformal sense.
A further limit of (ii) occurs when the singularity remains null for all time
and there is no trapped region.}
\label{collapse}
\end{figure}

Moreover, Christodoulou\cite{C} showed analytically that,
for the massless Einstein-Klein-Gordon system, 
there are non-generic collapses to a central null singularity
and non-generic black holes preceded by a central null singularity,
as depicted in Fig.\ref{collapse}.
This suggests the following picture of critical collapse:
as a family of black-hole solutions approaches criticality,
the trapped region is zipped up between the trapping horizon and the centre.
The closed zip is both a centre $r=0$ and, 
in a conformal sense, a trapping horizon $\partial_+r=0$,
and therefore is a central null singularity.

\noindent
Acknowledgements.
Thanks to Matt Choptuik, Andrei Frolov and Carsten Gundlach for discussions
and Abhay Ashtekar and the Center for Gravitational Physics and Geometry 
for hospitality.
Research supported by the National Science Foundation under award PHY-9800973.

\end{document}